\documentclass[final,3p,times,twocolumn,sort&compress]{elsarticle}
\usepackage{epic,eepic}

\usepackage{graphicx}
\usepackage{amssymb}
\usepackage{amsmath}

\usepackage{xspace}

\usepackage{courier}
\usepackage{helvet}
\usepackage[slantedGreek]{mathptmx}

\usepackage{lineno}
\usepackage[dvips,colorlinks=true,pagebackref=true]{hyperref}

\usepackage{booktabs}
\usepackage{threeparttable}

\newcommand{\degree}{^\circ}

\newcommand{\labr}{LaBr$_3$:Ce\xspace}
\newcommand{\higs}{HI$\gamma$S\xspace}
\newcommand{\gdrei}{$\gamma^{3}$\xspace}
\newcommand{\sulfur}{$^{32}$S\xspace}
\newcommand{\tgamma}{$\gamma$\xspace}
\newcommand{\gammagamma}{$\gamma$-$\gamma$\xspace}
\newcommand{\gammaray}{$\gamma$-ray\xspace}
\newcommand{\gammarays}{$\gamma$-rays\xspace}
\newcommand{\jone}{$J = 1$\xspace}
\newcommand{\etal}{\textit{et al.}\xspace}
\newcommand{\via}{via\xspace}

\hypersetup{pdftitle={The high-efficiency gamma-ray spectroscopy setup gamma3 at HIgS},
            pdfauthor={Bastian Loeher},
}

\begin{document}

\begin{frontmatter}


% \thanks[label1]{}^

\title{The high-efficiency \gammaray spectroscopy setup \gdrei at \higs}


\author[label1,label2]{B.~L\"oher\corref{cor}} \ead{b.loeher@gsi.de}
\author[label4]{V.~Derya}
\author[label5,label9]{T.~Aumann}
\author[label5]{J.~Beller}
\author[label6]{N.~Cooper}
\author[label5]{M.~Duch\^ene}
\author[label4]{J.~Endres}
\author[label1,label2]{E.~Fiori}
%\author[label4]{A.~Hennig}
\author[label1,label2]{J.~Isaak}
\author[label3,label8]{J.~Kelley}
\author[label5]{M.~Kn\"orzer}
%\author[label6]{F.~Naqvim}
\author[label5]{N.~Pietralla}
%\author[label3]{R.~Raut}
\author[label5]{C.~Romig}
%\author[label3]{G.~Rusev}
\author[label1,label2]{D.~Savran}
\author[label5]{M.~Scheck}
\author[label5]{H.~Scheit}
\author[label1,label2]{J.~Silva}
\author[label7]{A.~Tonchev}
\author[label3]{W.~Tornow}
\author[label3]{H.~Weller}
\author[label6]{V.~Werner}
\author[label4]{A.~Zilges}

\address[label1]{ExtreMe Matter Institute EMMI and Research Division, GSI Helmholtzzentrum f\"ur Schwerionenforschung, Planckstr. 1, 64291 Darmstadt, Germany}
\address[label3]{Department of Physics, Duke University and Triangle Universities Nuclear Laboratory, Durham, North Carolina 27708-0308, USA}
\address[label2]{Frankfurt Institute for Advanced Studies FIAS, Ruth-Moufang-Str. 1, 60438 Frankfurt am Main, Germany}
\address[label7]{Physics Division, Lawrence Livermore National Laboratory, Livermore, California 94551, USA}
\address[label8]{Department of Physics, North Carolina State University, Raleigh, North Carolina 27607, USA}
\address[label9]{GSI Helmholtzzentrum f\"ur Schwerionenforschung, Planckstr. 1, 64291 Darmstadt, Germany}
\address[label4]{Institut f\"ur Kernphysik, Universit\"at zu K\"oln, Z\"ulpicher Str. 77, D-50937 K\"oln, Germany}
\address[label5]{Institut f\"ur Kernphysik, TU Darmstadt, Schlossgartenstr.\ 9, 64289 Darmstadt, Germany}
\address[label6]{WNSL, Yale University, P.O. Box 208120, New Haven, CT 06520-8120, USA}

\cortext[cor]{Corresponding address: ExtreMe Matter Institute EMMI and Research Division, GSI Helmholtzzentrum f\"ur Schwerionenforschung, Planckstr. 1, 64291 Darmstadt, Germany, Tel.: +49 6159711803; fax: +49 6159713475}

\begin{abstract}

The existing Nuclear Resonance Fluorescence (NRF) setup at the \higs facility at the Triangle Universities Nuclear Laboratory at Duke University has been extended in order to perform \gammagamma coincidence experiments.
%This allows to investigate the decay behavior of photo-excited states with up to now unprecedented sensitivity. 
%The method of $\gamma$-$\gamma$ coincidence experiments has been used in the past with great success in particle-induced reactions in order to study in detail the decay pattern of nuclear excitations. 
%However in contrast to the NRF reaction, particle-induced reactions often do not (or only weakly) excite low-spin states, especially \jone states, at high excitation energy.
The new setup combines large volume \labr detectors and high resolution HPGe detectors in a very close geometry to offer high efficiency, high energy resolution as well as high count rate capabilities at the same time.
The combination of a highly efficient \gammaray spectroscopy setup with
%to perform $\gamma$-$\gamma$ coincidences
the mono-energetic high-intensity photon beam of \higs provides a worldwide unique experimental facility to investigate the \tgamma-decay pattern of dipole excitations in atomic nuclei.
%Several experiments will be carried out within a campaign, which make use of the high efficiency and $\gamma$-$\gamma$ coincidence capabilities.
%A feasibility test with a reduced setup has shown that the fully extended NRF setup provides the necessary performance to deliver new results in a range of topics in nuclear structure physics.
The performance of the new setup has been assessed by studying the nucleus \sulfur at 8.125~MeV beam energy.
The relative \tgamma-decay branching ratio from the $1^+$ level at 8125.4~keV to the first excited $2^+$ state was determined to 15.7(3)~\%.

\end{abstract}

\end{frontmatter}

% main text
\section{Introduction}\label{sec_introduction}
Structural features of excited nuclear states are reflected in their decay pattern and the partial decay widths $\Gamma_i$ of transitions to low-lying excited states or the ground state are directly linked to electromagnetic transition matrix elements.
%The decay pattern of excited states of the atomic nucleus is often connected to important details about the structure of a specific excitation.  
%The decay widths of transitions to individual low-lying excited states or the ground state are directly linked to the corresponding transition matrix elements.
Thus, the single decay channels are sensitive to different components in the wave function.  
This is especially true in the case of transitions to low-lying excited states.
For these levels the de-excitation takes place \via different components in the wave function compared to the excitation from the ground state. 
Therefore, the observation of these transitions and the determination of branching ratios reveals important experimental information needed to provide stringent and sensitive tests to modern model calculations.
The method of in-beam \gammaray coincidences in the spectroscopy of the $\gamma$-decay of excited states in combination with charged particle induced reactions has been proven to be a powerful tool for measuring even small branching ratios to excited low-lying states. 
Until now \gammagamma coincidence experiments have been performed mainly in combination with particle-induced reactions \cite{nola1982,cede1995}. 
In such reactions high-lying \jone states (in even-even nuclei) are often not or only very weakly excited, which strongly limits the investigation of the decay behavior of excitation modes such as the nuclear scissors mode \cite{bohl84b,loiu1993,heyd2010} or the Pygmy Dipole Resonance \cite{herz1997,paar07,Sav11,savr2013}. 
Consequently, for these excitation modes the decay behavior was not studied in detail so far. 
The reaction of choice for studying these modes is nuclear resonance fluorescence (NRF), which nearly exclusively populates \jone states \cite{knei1996}.
NRF performed with single \gammaray spectroscopy is primarily sensitive to $\Gamma_0 \frac{\Gamma_0}{\Gamma}$ with $\Gamma_0$ being the decay width to the ground state and $\Gamma$ the total width.
The observation of weakly branching transitions is difficult, especially in experiments using bremsstrahlung \cite{beli2001,schw2005,sonn2011}.
%whereas the \gammagamma coincidence method allows to directly measure $\Gamma_0 \frac{\Gamma_i}{\Gamma}$, and to determine $\Gamma_0$.
Therefore, the combination of an intense mono-energetic photon beam, which defines the excitation energy, with \gammagamma coincidence spectroscopy of the following decays offers ideal conditions to investigate in detail the decay behavior of photo-excited states.
First NRF experiments using the completely polarized \gammaray beam from a laser Compton backscattering (LCB) facility have been carried out by Ohgaki \etal \cite{ohga1994}
The first application of this method to investigate parities of nuclear levels was done by Pietralla \etal \cite{piet02a,piet02b}.

\begin{figure}[!t] \centering
\includegraphics[width=0.5\textwidth]{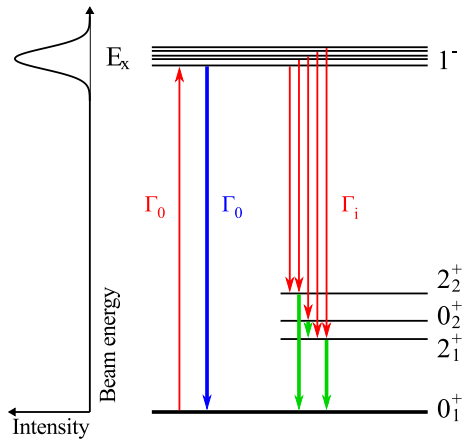}
\caption[Decay principle]{ After resonant excitation of a \jone state within the narrow energy profile of an incident photon beam, this state can
de-excite either directly to the ground state, or \via intermediate
states. Triggering on coincidences and gating on the secondary
transition allows to isolate the primary transitions.
}\label{fig_principle}
\end{figure}

The principle of the \gammagamma coincidence method in combination with a mono-energetic \gammaray beam is illustrated in Fig.~\ref{fig_principle}.  
After excitation by photo absorption the high-lying state at excitation energies $E_{x}$ may de-excite either directly to the ground state ($\Gamma_{0}$) or \via intermediate states ($\Gamma_{i}$). 
By detecting two emitted $\gamma$-rays stemming from a cascade in coincidence, even small branching ratios $\Gamma_{i}/\Gamma_{0}$ can be determined with good precision since non-resonant background produced by the photon beam in the target \via atomic processes is strongly suppressed.

This article reports on the installation and commissioning of a new experimental array, the \gdrei setup, at the High Intensity $\gamma$-ray Source (\higs \cite{scot96}).
This array provides sufficient efficiency in order to perform \gammagamma coincidence decay spectroscopy in combination with a high intensity, narrow bandwith photon beam.  
The design of the \gdrei setup is based on three principles: 
High efficiency, high energy resolution and high count rate capability.
A setup with a high full energy peak efficiency is mandatory to investigate small branching ratios with sufficient statistics within a reasonable amount of time. 
This condition requires maximum solid angle coverage which, using few detectors, implies a close geometry and demands high count rate capabilities.  
High energy resolution is needed when investigating higher lying states in energy regions, where the level density is large. 
A combination of High-Purity Germanium (HPGe) and LaBr detectors in a close geometry fulfills these requirements.

The newly installed experimental setup is described in Section~\ref{sec_setup}. 
Section~\ref{sec_daq} describes the different parts of the data acquisition hardware and software which are used for the \gdrei array. 
The details of the analysis and the results of a commissioning measurement on \sulfur are presented in Section~\ref{sec_analysis}.

\section{Experimental Setup}\label{sec_setup}
\begin{figure}[!t]
\centering
\includegraphics[width=0.5\textwidth]{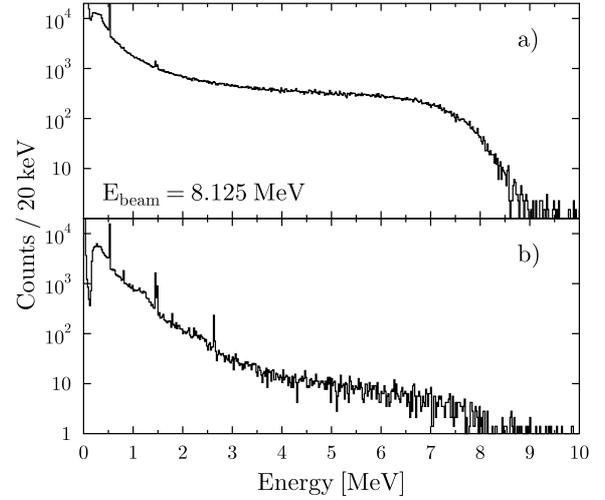}
\caption[Effect of evacuated beam pipe]{
Effect of evacuated beam pipe as seen by HPGe detectors. Both measurements were performed without target: a) Measurement with air. b) Beam pipe is evacuated. All other conditions are the same. Background contributions in the critical energy region from 2 to 9~MeV have been reduced by at least one order of magnitude.
}\label{fig_vacuum}
\end{figure}

The High Intensity $\gamma$-ray Source at the Triangle Universities Nuclear Laboratory (TUNL) at the Duke University is perfectly suited to perform NRF experiments on stable nuclei \cite{piet02b,well09}. 
\higs provides a high intensity ($> 50 $ $\gamma$/eV/s) polarized photon beam with a narrow energy spread ($\approx$3\%) and a variable energy (1 MeV to currently 100 MeV). 
The beam is produced \via laser Compton backscattering in the optical resonator of the storage ring based FEL.
The new \gdrei setup is located in the \higs upstream target room (UTR) which is situated about 57.2~m from the collision point.
The photon beam is first collimated to the appropriate diameter and then transported to the target through a plastic beam pipe with a wall thickness of about 3~mm, which can be evacuated.
This collimation is variable and results in beam diameters between 1.2 and 1.9~cm.
%If the target can withstand being placed in vacuum, it is recommended to evacuate the beam pipe, because otherwise a significant contribution from small angle scattering of photons on air is seen by the detectors.
A comparison of two background spectra taken with and without air in the photon beam pipe under otherwise identical conditions and with a mean beam energy of 8.125~MeV is shown in Fig.~\ref{fig_vacuum}, where this effect becomes apparent.
The contribution from atomic scattering on air to the observed background is substantial over the whole energy range up to the beam energy.
When the beam pipe is evacuated, this contribution is reduced by at least an order of magnitude in the energy region from 2 to 9~MeV.
The formerly invisible peaks stemming from natural background ($^{40}$K and $^{208}$Tl) emerge from the background.
The average count rate per LaBr (HPGe) detector is decreased from 1.4~kHz to 300~Hz (2.5~kHz to 250~Hz).
%Directly in front of the setup a 10" thick lead wall absorbs most of the beam halo radiation.
The previous NRF setup consisting of four 60\% HPGe detectors has been used for various photon scattering experiments (e.g. see Refs.~\cite{tonc10,piet09,isaa11}) proving the excellent performance of the \higs beam.
This setup is extended by additional detectors in order to provide sufficient efficiency for \gammagamma coincidence experiments.
%at the new \gdrei setup to investigate $J^{\pi}=1^-$ states in the energy range from 2 to 9 MeV.

\begin{figure}[!t]
\centering
\includegraphics[width=0.5\textwidth]{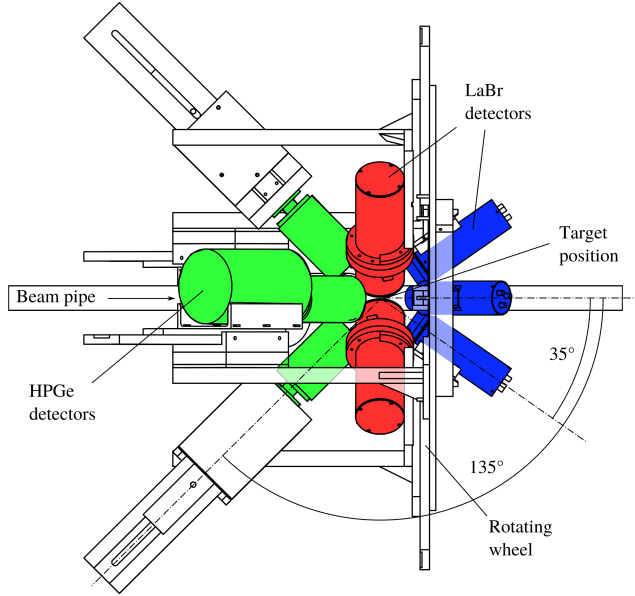}
\caption[Experimental setup]{
Schematic drawing of the \gdrei experimental setup. HPGe detectors are positioned at an angle of $\theta = 135\degree$ with respect to the beam pipe. The \labr detectors are used at $\theta = 90\degree$ and $\theta = 30\degree$ with respect to the beam pipe. The 3"x3" \labr detectors can also be exchanged with the HPGe detectors, and the wheel can be rotated by $\Delta \phi = 45\degree$ around the beam axis.
}\label{fig_schematic}
\end{figure}

Figure~\ref{fig_schematic} shows a schematic drawing of the newly installed \gdrei detector array.
It combines the four 60\% HPGe detectors with four 3"x3" \labr scintillators and four 1.5"x1.5" \labr scintillators.
All detectors are cylindrical in shape.
The 3"x3" \labr detectors are read out using photomultiplier tubes (PMT) of type R6233-100 SEL from Hamamatsu (3"x3") and voltage dividers produced in-house following a design from University Milan and INFN, Milan. 
Two of the 1.5"x1.5" detectors had PMTs of type R9779 from Hamamatsu (1.5"x1.5"), while the third one was read out using a Photonis XB2020.
The forth position was unoccupied during the experiment described in this article.
This solution with only few detectors allows for a flexible setup, where the high energy resolution is obtained from the HPGe detectors, while at the same time the \labr detectors positioned as close as possible to the target provide the necessary efficiency of at least a few percent (see Fig.~\ref{fig_efficiency} below).
The \labr detectors can be placed very close to the target (5.5 cm), because they can be operated at very high event rates \cite{loeh12}.
The basic layout uses the HPGe detectors in the backward position ($\theta = 135\degree$), and the large \labr detectors at an angle of $\theta = 90\degree$ with respect to the beam axis.
The smaller \labr detectors are placed at forward angle ($\theta = 30\degree$).
The layout is flexible, because the holding structures are interchangeable for the different positions, so that for specific experiments HPGe and \labr detectors can be exchanged.
All detectors are mounted on a wheel, that can be rotated up to $\Delta \phi = 45\degree$ around the beam axis, to position either the $\theta = 90\degree$ or the $\theta = 135\degree$ detectors in the (horizontal) polarization plane of the beam.
Additional mounts for the forward detectors allow them to be installed at an angle of $90\degree$ as well.
This increases the sensitivity when measuring angular distributions.
The detector mounts allow for a variable distance to the target, which is limited to a minimum of 5~cm for the $90\degree$ detectors and 8~cm for the $135\degree$ detectors.
The individual detectors are wrapped in a thin cylindrical layer of lead with 2~mm thickness to reduce scattering of reaction products from one detector into the active volume of the others.

To absorb low-energy $\gamma$-rays and reduce the count rate in the detectors, a set of Cu, Cd and Pb filters are available to be installed in front of the detectors.
The thickness of these filters can be adjusted to the conditions for each individual target and beam energy.
Additionally, a lead shield with 3.3~mm thickness can be placed around the plastic beam pipe.
%The standard layout recommends 0.5~mm Cu together with 2~mm of Pb for the \labr detectors and 1~mm of Cu together with 2~mm of Pb for the HPGe detectors.
%Depending on the actual target and the expected $\gamma$-ray energies it might be necessary to adjust these values.

The beam energy is measured between production runs with a 123\% HPGe detector ($\theta = 0\degree$) mounted on a remotely operated movable stand.
To decrease the beam intensity for the duration of this measurement, five different copper attenuators with thicknesses ranging from 1" to 3" can be inserted into the beam some 45~m upstream of the detector location.
This detector can also be used to determine the absolute photon flux.
For this purpose a 1 mm thick Cu plate is placed in the beam at a distance of 110~cm behind the target.
The elastically scattered photons from this plate are then detected by the HPGe detector at a specific angle ($7\degree$ in this case).
It is possible to derive the absolute photon flux from the kinematics.

\begin{figure}[!t]
\centering
\includegraphics[width=0.5\textwidth]{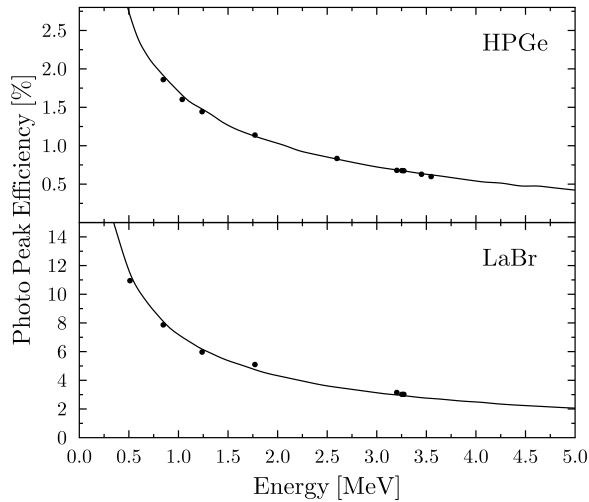}
\caption[Total full energy peak efficiency of the \gdrei setup]{
Total full energy peak efficiency of the \gdrei setup using the standard geometry shown in Fig.~\ref{fig_schematic}. Data points represent values measured with a calibration source ($^{56}$Co), smooth curve shows data from a Monte-Carlo simulation (GEANT4).
\textit{Top:} HPGe array.
\textit{Bottom:} \labr array.
}\label{fig_efficiency}
\end{figure}

The measured absolute full energy peak efficiency of the setup as a function of the $\gamma$-ray energy is shown in Fig.~\ref{fig_efficiency}.
Data points represent values measured with a calibration source ($^{56}$Co), while the smooth curve shows the results from a Monte-Carlo (MC) simulation done with GEANT4 \cite{alli06}.
For the given values the 3"x3" \labr detectors were placed at the $\theta = 90\degree$ position at a distance of 5.5~cm to the center of the beam, while the HPGe detectors were at an average distance of 9~cm.
Cu filters with a thickness of 1~mm have been used for the HPGe and 0.5~mm for the \labr detectors.
In the same configuration an average count rate of only 400~Hz per detector with full beam and without a target has been obtained.
The count rate induced only by natural background and the intrinsic radiation of the \labr detectors is on average 100~Hz per detector.
%The drop in efficiency below 1.0~MeV for \labr and 0.5~MeV for HPGe is due to the threshold of the trigger logic (see below) which has not been modeled in the simulation.
Exchanging the position of the HPGe with the large \labr detectors results in a decrease of the total efficiency of the setup by a factor of 4.

\section{Data Acquisition}\label{sec_daq}
Data acquisition is done using the GSI Multi Branch System (MBS) \cite{esse2000}, which operates on a VME single-slot PowerPC of type CES RIO4 and reads the data from the VME modules on an event by event basis to produce binary list mode data (LMD).
%Additionally the HPGe energy signals are converted by a 14~bit Canberra MCA and read out with the GENIE 2000 data acquisition system to produce single trigger histograms.
% The first output of each HPGe detector is passed through Canberra spectroscopy amplifiers, and then split into three signals, of which the first two are used in the timing branch, while the third one is converted by a 13~bit Mesytec MADC32 analog peak-sensing ADC.
% The secondary outputs from the HPGe detectors are directly connected to a 16~bit Struck SIS3302 flash ADC, running the specialized Gamma-Firmware (V1412).
% This firmware uses digital signal processing algorithms to determine the arrival time and energy of the pulses, and also adds information about pile-up and triggered channels.
% The \labr detector signals on the other hand are first amplified by timing filter amplifiers (TFA) and then also split into three signals each.
% Two of these signals are used in the timing branch, and the third signal is delayed by about 300~ns using passive delays and converted with a CAEN v965 QDC [REF].

The first output of each HPGe detector is split into two signals, which are used in the timing branch, while
the secondary outputs are directly connected to a 16~bit Struck SIS3302 flash ADC operated at a frequency of 100~MHz, with the specialized Gamma-Firmware (V1412) \cite{SIS3302}.
This firmware uses digital signal processing algorithms to determine the arrival time and energy of the pulses, and also adds information about pile-up and triggered channels.
The \labr detector signals on the other hand are first amplified by timing filter amplifiers (TFA) and then split into three signals each.
Two of these signals are used in the timing branch, and the third signal is delayed by about 300~ns using passive delays and converted with a CAEN v965 QDC.

The timing branch uses two constant fraction discriminators (CFD) for each detector.
Reasonable values for the thresholds have to be determined for each experiment individually.
This layout allows to construct trigger conditions that require both a \gammaray with a high energy and one with a lower energy in coincidence.
Using standard NIM electronics a number of low-level triggers are composed from the CFD outputs, such as OR triggers for each of the detector groups (HPGe, large \labr, small \labr) and multiplicity~2 triggers (i.e. two detectors of the same kind have triggered) for each group.
These triggers are then fed into the GSI Vulom4 VME Logic Module running the Trigger Logic II (TrLoII) firmware.
This firmware combines the incoming low-level triggers using coincidence or anti-coincidence conditions, and thus produces higher-level triggers, such as coincidence triggers between detector groups.
Any combination of the trigger inputs is possible to be selected and changed during operation.
Furthermore the TrLoII provides convenience features, such as a dead-time locking mechanism, scalers for the individual triggers, trigger reduction factors, gate generators and periodic pulses.
The use of the TrLoII also increases the setup's flexibility, for example it is possible to switch very quickly between different trigger schemes, e.g. for calibration measurements, beam energy measurements or production runs.
The \gdrei setup makes extensive use of these features and thus reduced the overall amount of NIM electronics necessary, resulting in a very compact 1-rack setup.
If a trigger is accepted by the TrLoII (i.e. the trigger arrived while the deadtime mechanism is unlocked), it produces a trigger decision within 45~ns and provides a main trigger signal to the VME modules.

A CAEN v775 TDC is used to record the detector times relative to the main trigger signal from the TrLoII, as well as the times of each generated low-level trigger.
To provide the count rates of the individual detectors for the low and the high threshold, a CAEN v830 scaler is used.
The scaler, the TDC and the Struck flash ADC directly receive the main trigger, while for the QDC matching gates are produced.
%Since the Master Start signal has an intrinsic jitter of 10~ns, in order to achieve the best energy resolution, the QDC gate is produced 
A logic AND of the main trigger and the properly delayed OR trigger of the \labr detectors, which are aligned to the delayed energy signals, is produced.
%For accurate dead-time measurements a pulser with a frequency of 100~kHz occupies a separate trigger input.
%A large downscale factor is applied, such that only about 100 events per second are actually recorded.
%These events can additionally be used to determine the QDC pedestal value.
% \begin{figure}[!t]
% \centering
% \includegraphics[width=0.5\textwidth]{fig/rf.eps}
% \caption[Time difference between detector and beam pickup]{
% Time difference between detector and beam pickup.
% }\label{fig_rf}
% \end{figure}
A beam pickup signal from the electron storage ring of the DFELL is also connected to one of the trigger inputs and its arrival time is recorded by the TDC.
This time is correlated with the production of a burst of $\gamma$-rays from the FEL,
which allows to measure the time difference between the reaction trigger and the photon beam, in order to discriminate uncorrelated background events.
Furthermore it prevents event mixing from adjacent beam bursts.
% Fig.~\ref{fig_rf} shows this time difference histogram for \labr detector 1.
% A large coincidence peak can be seen in the middle, surrounded by background events.

The recorded data is directly available as a data stream from the MBS, and in parallel is unpacked to a ROOT tree using the \textit{ucesb} unpacker \cite{joha10}.
For monitoring purposes during runtime of the experiment, a histogramming tool (GHOST, Gamma Histogramming and Online Spectra Tool) has been constructed based on the ROOT analysis framework \cite{Roo10}.
It reads concurrently from the ROOT tree and allows to monitor calibrated energy and time histograms for all detectors, view coincidence matrices, summed histograms and includes the possibility to apply cuts to the data already at this early stage.
%This enables the user to make decisions for setup changes or parameter adjustments based on data available in real-time.
The modular and decoupled layout of the DAQ, unpacking and histogramming parts increases the system's stability.

\begin{figure}[!t]
\centering
\includegraphics[width=0.5\textwidth]{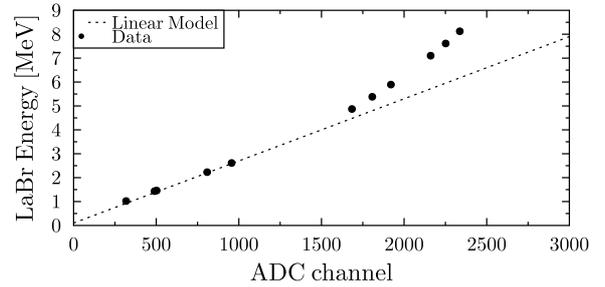}
\caption[LaBr non-linearity]{
Non-linear behavior of LaBr. A linear model fitted up to an energy of 2.614~MeV is shown together with the actual data points. Uncertainties range within the size of the markers.
}\label{fig_nonlinear}
\end{figure}

\begin{figure}[!t]
\centering
\includegraphics[width=0.5\textwidth]{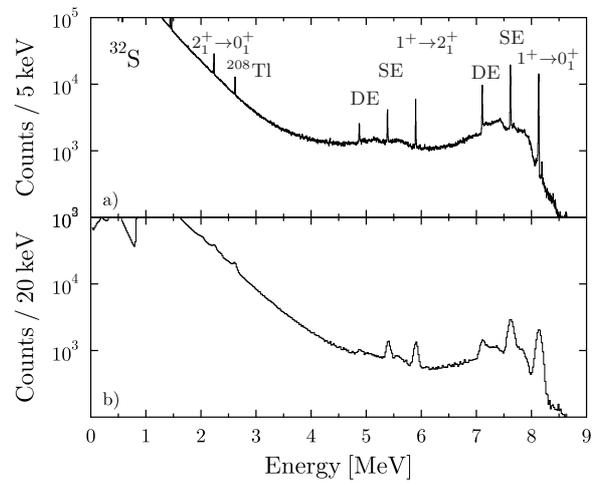}
\caption[Raw spectra]{
Single trigger spectra (summed over all detectors of each type) for \sulfur at a beam energy of 8.125~MeV. 
\textit{Top:} HPGe
\textit{Bottom:} \labr.

}\label{fig_raw}
\end{figure}

\begin{figure}[!t]
\centering
\includegraphics[width=0.3\textwidth]{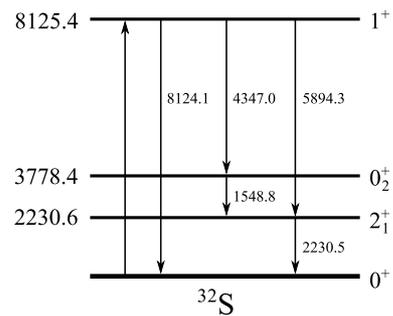}
\caption[Level scheme]{
Simplified level scheme of \sulfur, showing the important transitions which are discussed. Energies in keV.
}\label{fig_levels}
\end{figure}

\section{Data Analysis and Results}\label{sec_analysis}
In the commissioning beam time for the \gdrei setup the nucleus \sulfur was investigated in a 4~h experimental run.
This section gives an overview over the important steps in the data analysis and presents the results from this commissioning experiment.

The analysis of the experimental data recorded at the \gdrei setup includes basic steps such as energy and efficiency calibration using $\gamma$-ray sources with precisely known activity, and some additional steps unique to the \higs facility.
The energy calibration of the \labr detectors has to be done using a cubic polynomial, since the response of these detectors is noticeably non-linear at energies above 2 MeV  as shown in Fig.~\ref{fig_nonlinear} for the known transitions in \sulfur and \gammarays from $^{56}$Co.
This effect is due to saturation effects in the PMT and could be alleviated by choosing a lower bias voltage at the cost of resolution at low energy.
For this particular analysis the non-linear behavior is not a severe problem and was taken into account.
The histograms corresponding to \labr and HPGe singles spectra are shown in Fig.~\ref{fig_raw}. 
Besides the peaks stemming from the known transition energies in \sulfur (see Fig.~\ref{fig_levels}), these spectra without any coincidence condition are dominated by a strong continuous background, which increases exponentially towards lower \gammaray energies.

\begin{figure}[!t]
\centering
\includegraphics[width=0.5\textwidth]{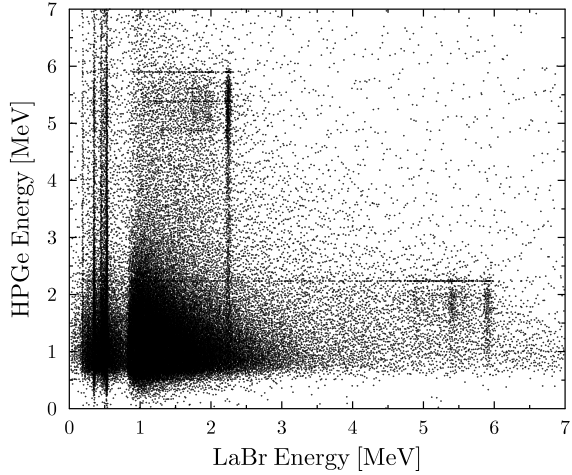}
\caption[Coincidence Matrix HPGe-LaBr]{
The \gammagamma coincidence matrix for HPGe-LaBr coincidences.
At gamma-ray energies of simultaneously emitted photons intensity maxima show up accompanied by horizontal and vertical lines, due to the detector response.
}\label{fig_coinc2}
\end{figure}

\begin{figure}[!t]
\centering
\includegraphics[width=0.5\textwidth]{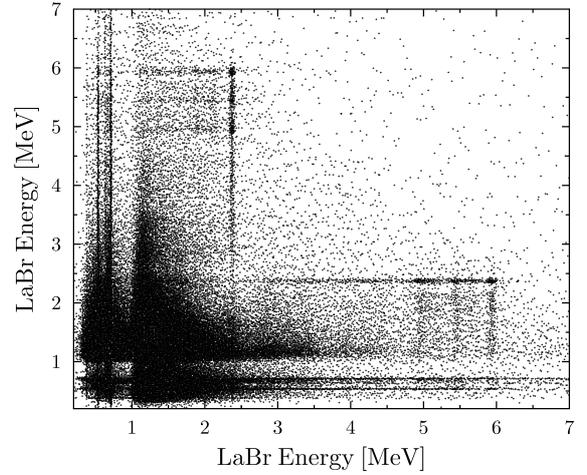}
\caption[Coincidence Matrix LaBr-LaBr]{
Same as Fig.~\ref{fig_coinc2}, but for LaBr-LaBr coincidences.
}\label{fig_coinc}
\end{figure}

\begin{figure}[!t]
\centering
\includegraphics[width=0.5\textwidth]{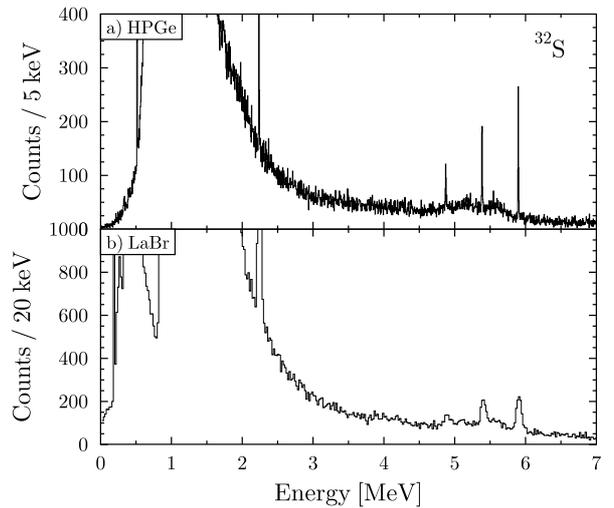}
\caption[Matrix projection]{
Projection of the HPGe-LaBr matrix (Fig.\ref{fig_coinc2}) on each of the axes in the energy region from 0 to 7~MeV.
}\label{fig_proj}
\end{figure}

Combining any two energies measured in each event in the coincidence data sub-set, a two-dimensional \gammagamma coincidence matrix can be produced either for HPGe-LaBr or LaBr-LaBr coincidences.
The matrices built from the \sulfur data are shown in Fig.~\ref{fig_coinc2} and Fig.~\ref{fig_coinc}, respectively.
The coincidence time window for any two detectors was set to 60~ns for the HPGe-LaBr case and 20~ns for the LaBr-LaBr case in the time spectra.
The resulting data do not depend strongly on variations of this time window, because the condition on the time structure of the pulsed beam already discriminates uncorrelated background events.
The coincident events stemming from a \gammaray cascade appear as vertical or horizontal lines in the matrix, which represent the detector response of the specific detectors, comprised of the full energy peak (FEP), single escape (SE) and double escape (DE) peaks as well as the Compton background.
Figure~\ref{fig_proj} shows the projection of the HPGe-LaBr matrix on each of the two axes.
In these spectra the FEP, the SE and the DE peaks of the $1^+ \rightarrow 2^+_1$ transition in \sulfur are clearly visible and the underlying background is reduced compared to the singles spectra (Fig.\ref{fig_raw}).
The full strength of the \gammagamma coincidence method becomes apparent when using the energy information of both detected photons.
Applying the condition that one \labr detector has detected a \gammaray corresponding to the energy of the $2^+_1$ state in \sulfur, allows to project a nearly background free energy spectrum in the HPGe detector, showing exclusively the feeding transitions (plus corresponding detector response) to the $2^+_1$ state (Fig.~\ref{fig_32s}(a)).

%With the polarized photon beam it is also possible to study the multipolarity of the transitions and assign parities to the measured states.
%This has already been done quite frequently, as described in \cite{piet02b,fran04,savr05}.

\subsection{Performance Study}\label{sec_performance}
%In a two day commissioning beam time for the new \gdrei setup the goal was to optimize detector shielding and setup parameters and to show that this new setup is capable of delivering the desired performance.
As a test case the isotope \sulfur was chosen for two reasons: The level scheme of this nucleus is well investigated and includes a $J^{\pi} = 1^+$ state at 8.125~MeV, which is strongly excited in the (\tgamma,\tgamma') reaction.
In addition this state decays into the $2_1^+$ state with a considerable branching.
A simplified level scheme with the relevant information of the $1^+$ state is shown in Fig.~\ref{fig_levels}.

%With the beam energy of 8.125~MeV the $1^+$ state in \sulfur at 8.125~MeV was excited and (as illustrated in the simplified level scheme in Figure~\ref{fig_levels}) the properties of both the elastic decay back to the ground state as well as the inelastic decay through the $2^+_1$ state are known \cite{endt90,babi02b}.
The branching ratio for the transition $1^+ \rightarrow 2^+_1$ was determined in \cite{babi02b} to be 16(4)\%. %14(3)\%.
Other decay branchings were previously unknown.
Within only 4~hours of beam time enough statistics were accumulated in all transitions in order to allow for a detailed analysis.

Using a 5.2~g natural sulfur target (\sulfur abundance: 95\%) enclosed in a polyethylene container the count rates in the individual 3"x3" \labr detectors were about 75~kHz and in the HPGe detectors about 10~kHz.
The distance to the target was 5.5~cm for the \labr and between 8 and 9~cm for the HPGe detectors.

\begin{figure}[!t]
\centering
\includegraphics[width=0.5\textwidth]{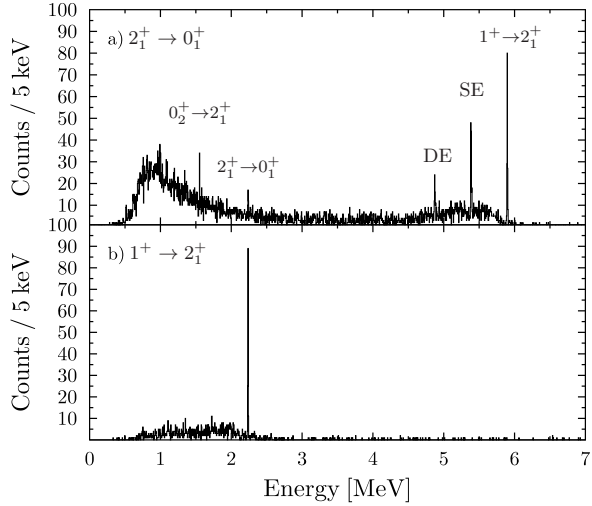}
\caption[Matrix projection]{
Coincidence spectra of the HPGe array: a) Gated on transition $2^+_1 \rightarrow 0^+_1$ at an energy of 2.230~MeV measured in \labr.
b) Gated on transition $1^+ \rightarrow 2^+_1$ with an energy of 5.894~MeV.
}\label{fig_32s}
\end{figure}

In Fig.~\ref{fig_32s} the HPGe spectra are shown for two different gating conditions on the measured energies in the \labr detector array.
The upper part shows the spectrum with the energy gate on the transition $2^+_1 \rightarrow 0^+_1$ at 2.230~MeV in any \labr detector, while the lower part shows the same data with the cut on the energy of the primary transition $1^+ \rightarrow 2^+_1$ at 5.894~MeV.
The width of the gate was set to 100~keV for the low energy cut and 160~keV for the high energy cut to account for the resolution at the corresponding energies.

Compared to the singles spectrum shown in Fig.~\ref{fig_raw} the atomic background is strongly supressed in these HPGe spectra using the energy information of the \labr array.
The spectrum shown in Fig.~\ref{fig_32s}(a) is dominated by the transition $1^+ \rightarrow 2^+_1$ and its corresponding detector response.
In addition the peak stemming from the transition $0^+_2 \rightarrow 2^+_1$ is clearly visible, which is not the case in the singles data.
This shows the improved sensitivity to weak transitions using the coincidences.
Unfortunately the primary transition $1^+ \rightarrow 0^+_2$ could not be observed, since the HPGe detector efficiency at 4347 keV is too low.
The small remnant of the $2^+_1 \rightarrow 0^+_1$ transition observed in the spectrum is due to the background contributions in the LaBr energy cut.
No significant background counts are observed in the region between 6 and 8.2~MeV.
The histogram in the lower part of Fig.~\ref{fig_32s} shows as expected only the transition from the first excited state to the ground state.
No background or random coincidences have been subtracted.

%The energy-gated histogram in Fig.~\ref{fig_32s}(a) contains mainly the detector response to the photons from the inelastic transition $1^+ \rightarrow 2^+_1$, and a number of additional transitions, most notably the transition $0^+_2 \rightarrow 2^+_1$.
%At the gated energy of 2.230~MeV ($1^+ \rightarrow 2^+1$) a peak is still visible, because the energy gate also includes the background contributions below the peak (see Fig.~\ref{fig_proj}).
%The histogram in the lower part of Fig.~\ref{fig_32s} shows only the transition from the first excited state to the ground state, as expected.

The peak-to-background ratio (PBR) for the peaks of interest has been determined.
Both the peak and the background have been integrated in a 3$\sigma$ region around the peak.
A comparison of the PBR for the peaks at 2.230~MeV and 5.894~MeV is done to show the increase in sensitivity when applying the coincidence condition and the energy gate.
The PBR at 5.894~MeV increased from 1.03(2) in the HPGe singles spectrum to 11.7(13) using the coincidence to the \labr detectors.
In the LaBr-LaBr coincidence, the PBR increases from 0.45(1) to 4.3(4).
%, so roughly by a factor of 10.
At lower energies, where the spectrum is dominated by contributions from atomic background, the increase in PBR is even larger.
This is estimated by applying an inverse energy gate, thus requiring an energy of 5.894~MeV in the \labr detector, and comparing the PBR of the peak at 2.230~MeV.
The result is an increase of the PBR from 0.197(3) by a factor of roughly 50 to 9.6(15) in the case of HPGe-LaBr coincidences and an increase from 0.0506(5) to 10.2(12) for the LaBr-LaBr case.
This shows that the \gammagamma coincidence method increases the sensitivity of the setup by at least an order of magnitude. 
This is particularly noticeable in the region where usually the background from atomic scattering is dominant.

In the following the determination of a new value for the branching ratio $b_0$ of the transition $1^+ \rightarrow 2^+_1$ is described.
%This value could be determined once using HPGe singles data and once using HPGe-HPGe coincidences.
In the analysis the areas of the single peaks corresponding to the transitions were determined using a fit with Gaussian shape on a linear background.
These areas were then corrected for the energy dependent full energy peak efficiency of the respective detector.
The shape of the efficiency curves have been determined using MC simulations up to an energy of 10~MeV taking into account the specific geometry of the \sulfur measurement.
The accuracy of these shapes could be veryfied up to an energy of 3.6~MeV using measurements with radioactive sources.
Efficiency values for higher energies were taken directly from the MC simulation.
As a measure for the branching ratio $b_0$ the intensity of the transition $1^+ \rightarrow 2^+_1$ relative to the strongest transition $1^+ \rightarrow 0^+_1$ was determined.
%And then the value for $b_0 = \Gamma_i/\Gamma$ was derived by taking into account only these two transitions for $\Gamma$.

Exploiting the HPGe-LaBr coincidence data, while requiring the primary transition (5894~keV) to be detected in one of the \labr detectors, a new value of %0.142(10)
16.5(14)\% for the branching ratio $b_0$ of the transition $1^+ \rightarrow 2^+_1$ could be determined.
%For the case of the inverse LaBr-HPGe coincidence (primary transition detected in HPGe detectors), a value of 0.14(30) has been determined.
%The error is larger in this case because of the lower efficiency of the HPGe detectors at higher energy.
Since in this particular case the branching ratio is sufficiently large, it was also possible to determine the same value from the HPGe spectra without coincidence (Fig.~\ref{fig_raw}).
This leads to a value of %0.139(4).
15.7(3)\%.
The precision is higher compared to the value from coincidence data, due to the lower count rate in the coincidence data.
However, it should be stressed that in cases of weaker transitions a determination of branching ratios will become possible in the coincidence data only due to the much better PBR.
In the singles data weak transitions are hidden in the background.
Additionally LaBr-LaBr coincidences have been investigated.
The same conditions as for the HPGe-LaBr case have been applied.
The resulting value for the branching ratio of the transition $1^+ \rightarrow 2^+_1$ is % 0.136(5).
15.7(6)\%.
From the \labr spectra without the coincidence a value of %0.136(1)
15.8(1)\% could be determined.

The given errors are of pure statistical nature and do not account for uncertainties stemming from the efficiency calibration.
However, for the computation of the value derived from singles data only relative efficiencies are used.
As shown in \cite{cars2010}, the uncertainty introduced when deriving relative efficiencies from Geant4 simulations is about 2\%.
For the coincidence measurement on the other hand, absolute efficiencies for the two detector systems have to be compared.
These absolute efficiencies could be reproduced up to an energy of 3.6~MeV to within 1\% in comparison to measurements with precision calibration sources in an identical setup.
This leads to a total systematic error of 3\% for the coincidence measurement.

The results are summarized in Table~\ref{tab_br} and compared to data from \cite{babi02b} and the evaluated nuclear structure data file (ENSDF) \cite{viit1974}.

\begin{table}[h]
\centering
\begin{threeparttable}[b]
\resizebox{\columnwidth}{!}{%
\small\addtolength{\tabcolsep}{-3pt}
% \begin{tabular}{lcccccc} 
% \toprule
% $E_{\gamma}$ & \multicolumn{4}{c}{$b_0$ [\%]}          & $b_{0,a}$  & $b_{0,b}$\\ 
% \cmidrule(r){2-5}
% [keV]        & HPGe-LaBr\tnote{1} & LaBr-LaBr\tnote{2} & HPGe\tnote{3} & LaBr\tnote{4} &  [\%]  &   [\%]     \\
% \midrule
% 8124        & 85.8(10)  & 86.4(5)   & 86.1(4) & 86.3(1) & 86(19) & 85(5) \\
% 5894        & 14.2(10)  & 13.6(5)   & 13.9(4) & 13.6(1) & 14(3)  & 15(5) \\
% \bottomrule
% \end{tabular}%
\begin{tabular}{lcccccc} 
\toprule
$E_{\gamma}$ & \multicolumn{4}{c}{$b_0$ [\%]}          & $b_{0,a}$  & $b_{0,b}$\\ 
\cmidrule(r){2-5}
[keV]        & HPGe-LaBr\tnote{1} & LaBr-LaBr\tnote{2} & HPGe\tnote{3} & LaBr\tnote{4} &  [\%]  &   [\%]     \\
\midrule
8124        & 100      & 100     & 100     & 100         & 100   & 100  \\
5894        & 16.5(14) & 15.7(6) & 15.7(3) & 15.8(1)     & 16(4) & 18(6) \\
\bottomrule
\end{tabular}%
}
\begin{tablenotes}
{\scriptsize
\item[1] Value from coincidences between HPGe and \labr detectors
\item[2] Value from coincidences between \labr detectors
\item[3] Value from HPGe singles
\item[4] Value from \labr singles
}
\end{tablenotes}
\end{threeparttable}
\caption[$^{32}$S branching ratios]{Presently determined branching ratios $b_0$ in $^{32}$S. 
Compared to values $b_{0,a}$ from \cite{babi02b} and $b_{0,b}$ from ENSDF \cite{viit1974}. Only statistical errors are given. Systematic errors are on the 3\% level.}
\label{tab_br}
\end{table} 

\section{Conclusions and Outlook}\label{sec_conclusion}

In this article the new high-efficiency \gammagamma coincidence setup (\gdrei) at \higs has been introduced, and the results from the commissioning beam time have been presented.
The superior sensitivity to \gammarays resulting from the de-excitation of the nuclues \via intermediate states in the energy region of interest compared to single-\tgamma spectroscopy was verified on the test case of \sulfur.
The commissioning phase also included the investigation of the influence of an evacuated beam pipe and data acquisition parameters.
With the knowledge gained during this beam time the \gdrei setup was optimized to yield maximum performance during the following experiments.
The first experimental campaign covers the investigation of two-phonon states, the M1 scissors mode, as well as the decay pattern of the Pygmy Dipole Resonance in various atomic nuclei.
The feasibility of the \gammagamma coincidence method in combination with a mono-energetic photon beam has been successfully demonstrated at \higs.
Future facilities, such as the new ELI-NP experimental site featuring a very intense ($10^{13}\gamma/s$) brilliant photon beam and an unprecedented bandwidth of 0.1\% \cite{elin2013}, are perfectly suited for installing a \gammagamma coincidence setup similar to the \gdrei array.

\section*{Acknowledgments} 
The authors thank the TUNL and \higs technical administration and staff and the mechanical workshop for their great help in setting up the \gdrei system.
The work described in this article is supported by the Alliance Program of the Helmholtz Association (HA216/EMMI), the DFG (SFB~634 and ZI 510/4-2) and U.S.~DOE grants no. DE-FG02-91ER-40609 and no. DE-FG02-97ER-41033.

% \appendix
% \section{}
% \label{}

%\bibliographystyle{../../../bibtex/apsrev}
%\bibliography{../../../bibtex/english}
\bibliographystyle{apsrev}

\end{document}